\documentclass{jetpl}
\twocolumn

\lat

\title{Generalized unimodular gravity in Friedmann and Kantowski-Sachs universes}

\rtitle{Generalized unimodular gravity Friedmann and Kantowski-Sachs universes}

\sodtitle{Generalized unimodular gravity Friedmann and Kantowski-Sachs universes}

\author{A.\,Yu.\,Kamenshchik$^{+*}$\thanks{e-mail: kamenshchik@bo.infn.it},
A.\,Tronconi$^+$, G.\, Venturi$^+$}

\rauthor{A.\,Yu.\,Kamenshchik, A. Tronconi, G. Venturi}

\sodauthor{Kamenshchik, Tronconi, Venturi}

\address{
$^+$ Dipartimento di Fisica e Astronomia, Universit\`a di Bologna and INFN, via Irnerio 46, 40126 Bologna, Italy\\
$^*$L.D.Landau Institute for Theoretical Physics RAS,
117940 Moscow, Russia}

\dates{}{}

\abstract{
We illustrate the  recently proposed generalized unimodular gravity using simple examples of the Friedmann, Kantowski-Sachs and Schwarzschild geometries and show that it  can be further generalized and reveal some unexpected and interesting effects. 
}


\begin{document}

\maketitle

One of the oldest modified theories of gravity is  unimodular gravity, dating to the paper by Einstein \cite{Einstein}. 
The recent rebirth of this idea is connected with the papers \cite{Unruh,Henneaux}. The main point of  unimodular gravity consists of the fact that when one requires that the determinant of the metric is fixed,  the cosmological constant arises as an integration constant in the Einstein equations. The idea can be presented in a very simple way. To  limit ourselves to the variations of the metric which do not change its determinant (which are called volume-preserving diffeomorphisms):
\begin{equation*}
\delta g = gg^{ij}\delta g_{ij} = 0 \Rightarrow g^{ij}\delta g_{ij} = 0,
\end{equation*}
one adds to the Hilbert-Einstein action a term with a Lagrange multiplier: 
\begin{equation}
\int dx[ \sqrt{-g}(R-L_{\rm matter})-\lambda(\sqrt{-g}-1)].
\end{equation}
The variation of the action with respect to the metric then gives
\begin{equation}
R_{ij}-\frac12g_{ij}R-\lambda g_{ij} = T_{ij},
\end{equation}
where the Lagrange multiplier $\lambda$ plays the role of the cosmological constant.
It then follows immediately from the Bianchi identities that $\lambda$ is indeed a constant.
The unimodular gravity theories can be essentially generalized if one uses the Arnowitt-Deser-Misner (ADM) \cite{ADM} approach to gravity.
Let us represent the metric as 
\begin{eqnarray}
&&ds^2 = -(N^2-N_{\alpha}N^{\alpha})d\tau^2 + 2N_{\alpha}d\tau dx^{\alpha} +\gamma_{\alpha\beta}dx^{\alpha}dx^{\beta},\nonumber \\
&& \alpha=1,2,3
\label{ADM}
\end{eqnarray}
where $N$ and $N_{\alpha}$ are lapse and shift functions. 
Then 
\begin{equation}
-g = -\det g_{ij} = N^2 \gamma,\ \ \gamma = \det \gamma_{\alpha\beta}.
\label{det}
\end{equation}
In the theories of  unimodular gravity  one fixes the gauge condition as
\begin{equation}
g = -1,
\label{det1}
\end{equation}
which is equivalent to 
\begin{equation}
N = \frac{1}{\sqrt{\gamma}}.
 \label{unimod}
 \end{equation}
The variation of the action with respect to the lapse function gives the super-Hamiltonian constraint, which is equivalent to the $00$ component of the Einstein 
 equations. If we fix the lapse function before the variation, then the super-Hamiltonian constraint does not exist. However, from the other components of the Einstein equations, one can derive a first integral. This first integral is equivalent to the super-Hamiltonian constraint, but in this constraint an additional term arises. This term 
 represents a cosmological constant, which can have an arbitrary value. 
 
 The generalized unimodular gravity theory was suggested recently in the paper \cite{we}. Let us suppose that we have fixed the gauge as 
 \begin{equation}
 (-g^{00})^{-1/2} = N(\gamma),
 \label{fix}
 \end{equation}
 where $N$ is some function of the determinant of the three-dimensional metric. The corresponding action (in the absence of matter) can be written as 
\begin{equation}
\int dx[ \sqrt{-g}R-\lambda((-g^{00})^{-1/2} - N(\gamma))].
\end{equation}
 On varying this action with respect to the contravariant metric $g^{ij}$, we obtain the modified Einstein equations
 \begin{equation}
 R_{ij} -\frac12g_{ij}R = (\varepsilon + p)u_iu_j + g_{ij}p,
 \label{modified}
 \end{equation}
 where  in the right-hand side one has the energy momentum tensor of an effective fluid with the energy density 
 $$
 \varepsilon = \frac12\frac{\lambda}{\sqrt{\gamma}},
 $$
 and the pressure $p$ satisfies the equation of state
 $$
 p = w\varepsilon,
 $$ 
 where the equation of state parameter $w$ depends on the function $N(\gamma)$ as follows:
 \begin{equation}
 w = 2\frac{d\ln N(\gamma)}{d\ln \gamma}.
 \label{eq-state}
 \end{equation}
 Further, the four-velocity is
 $$
 u_i = -N\delta_i^0.
 $$
 One sees that by choosing in a proper way the function $N(\gamma)$ we can obtain in the model under consideration different types of effective fluids filling the universe.  
 It is important to observe that the energy density of this effective fluid can be negative as  well as positive. Let us note that in the context of dark energy models,  fluids with negative energy density were considered, for example, in papers \cite{Khal,Trans,late} for the Chaplygin gas \cite{Chap} and in paper \cite{string} for the string gas.

 Thus, on treating one of the Lagrange multipliers of the General Relativity, i.e. the lapse function $N$ not as a Lagrange 
 multiplier, but as a given function of other variables, we freeze one of the symmetries of the system and as a result the 
 effective matter content of the theory becomes richer.
 Let us note that this phenomenon is quite well-known and was pioneered by Dirac in the paper \cite{Dirac} dedicated to electrodynamics. Indeed, he choses
 a gauge  by expressing the temporal component of the electromagnetic potential, which plays the role of a Lagrange multiplier in electrodynamics, as a function of its spatial components. What we have done in the paper \cite{we}  is very similar to the Dirac's suggestion.  
Indeed, we  fix the lapse function as a function of the determinant of the spatial metric. In both  cases, additional kinds of matter arise in the theories, which initially consisted only of a pure electromagnetic field or a pure gravity.
  Let us give an exact citation from the paper \cite{Dirac}:
``In the theory of the electromagnetic field without charges, the potentials are not fixed by the field, but are subject to gauge transformations. The theory thus involves more dynamical variables than are physically needed. It is possible by destroying the gauge transformations to make the superfluous variables acquire a physical significance and describe electric charges.'' We can add that the Dirac's paper \cite{Dirac} has attracted the attention of researchers for a long time (see e.g. \cite{post0,post,post1,post2,post3}).  
A similar class of problems with the so called unfree gauge symmetry was considered recently in papers \cite{Semen,Semen1,Semen2}.

We wish to here add that some kind of effective matter also arises in different models suggested recently. Thus, in paper
\cite{Lim}, the authors  introduce a pair of  scalar fields, one of which  plays the role of a Lagrange multiplier, while the corresponding non-holonomous constraint makes the kinetic energy of the other scalar field always equal to its potential energy.
As a result in the universe an effective matter is present which, under  some conditions, can play the role of  a unified description for dark matter and dark energy. As in our approach, in paper \cite{Lim} the study of the constrained dynamics is essential. However, in contrast with  generalized unimodular gravity,  the gravitational degrees of freedom are uneffected in the model \cite{Lim}. We wish  to also mention  the  paper \cite{Mukhanov}, wherein  mimetic dark matter matter arises owing  to some interplay between the auxiliary metric and a scalar field.

In spite of its simplicity the model of  generalized unimodular gravity \cite{we} imposes some interesting problems and opens some attractive prospects due to its unexpected flexibility. In paper \cite{Barv} the Hamiltonian formalism for this model, treated as a rather complicated example of a constrained dynamical system \cite{constrained}, was considered in detail. Especially interesting in this context is the question of the determination of the number and the character of the physical degrees of freedom, arising here.
The paper \cite{Barv1} was devoted to the inflationary model based on  generalized unimodular gravity and the behaviour of linear perturbations in this model was studied.

However, the model \cite{we} opens some interesting opportunities already at the level of a simple minisuperspace models with finite number of degrees of freedom.   We shall discuss here some of them.
  
Let us begin with a flat Friedmann model with the metric 
\begin{equation}
ds^2= - N^2(t) dt^2 + a^2(t)dl^2.
\label{Fried}
\end{equation}
In this case 
$$
\gamma = a^6
$$
and the equation (\ref{eq-state}) is simply 
\begin{equation}
w = \frac13\frac{d\ln N(a)}{d\ln a}.
\label{eq-state1}
\end{equation}
We feel that it is rather instructive to derive Eq. (\ref{eq-state1}) directly from the Friedmann model.
The Lagrangian for the flat Friedmann universe without matter can be written as 
\begin{equation}
L = \frac{\dot{a}^2a}{N}.
\label{Lagrange}
\end{equation}
If we now  treat the lapse function as a function of the scale factor $a$,  the variation with respect to $a$ gives the following Euler-Lagrange equation:
\begin{equation}
2\frac{\ddot{a}a}{N}+\dot{a}^2\frac{d(a/N)}{da}=0,
\label{E-L}
\end{equation}
where the ``dot'' signifies the differentiation with respect to the time parameter $t$. 
This equation can be rewritten as 
\begin{equation}
\frac{1}{\dot{a}}\frac{d}{dt}\left(\frac{\dot{a}^2a}{N}\right) = 0.
\label{E-L1}
\end{equation}
which  means that now have the first integral
\begin{equation}
\frac{\dot{a}^2a}{N} = C,
\label{first}
\end{equation}
where $C$ is a constant. 
If now we divide Eq. (\ref{first}) by $Na^3$, we obtain 
\begin{equation}
\frac{\dot{a}^2}{N^2a^2} = \frac{1}{a^2}\left(\frac{da}{d\tau}\right)^2 = \frac{C}{Na^3},
\label{first1}
\end{equation}
where $\tau$ is the cosmic or synchronous  time 
$$
d\tau = Ndt.
$$  

The equation (\ref{first1}) can be interpreted as the first Friedmann equation for a flat universe filled with matter having the energy density
\begin{equation}
\varepsilon = \frac{C}{Na^3}.
\label{en-den-F}
\end{equation}
On remembering the energy conservation law 
\begin{equation}
\frac{d\varepsilon}{da} = -3\frac{\varepsilon+p}{a},
\label{conserv}
\end{equation}
we can immediately find the pressure 
\begin{equation}
p = -\frac13a\frac{d\varepsilon}{da}-\varepsilon=\frac{C}{3N^2a^2}\frac{dN}{da} = \frac13\frac{d\ln N}{d\ln a}\varepsilon,
\label{pressure}
\end{equation}
which confirms the relation (\ref{eq-state1}). 

It is now very easy to find the functions $N(a)$ corresponding to the required types of the energy density in the Friedmann models.
Thus, if one considers the universe filled with a  perfect fluid with the constant equation of state parameter $w = w_0$, then the energy density is 
\begin{equation}
\varepsilon = \frac{C}{a^{3(1+w_0)}}
\label{pres}
\end{equation}
and on comparing Eq. (\ref{pres}) and Eq. (\ref{en-den-F}), one finds (up to an arbitrary constant) 
\begin{equation}
N = a^{3w_0},
\label{N} 
\end{equation}
which for $N = 1/a^3$ gives the standard unimodular gravity, for $N = 1$ gives a universe filled with an effective dust etc. If one wishes to consider some more complicated models like, for example, the Chaplygin gas \cite{Chap} with the equation of state 
$$
p = -\frac{A}{\varepsilon},
$$ 
where $A$ is a positive and the energy density behaves as 
\begin{equation}
\varepsilon = \sqrt{A+\frac{B}{a^6}},\ B > 0,
\label{Chap}
\end{equation}
which is also possible. Indeed, in this case 
\begin{equation}
N = \frac{1}{\sqrt{Aa^6+B}}.
\label{Chap1}
\end{equation}

Let us now consider the question of a possible phantom divide line crossing. It is known that the observed cosmic acceleration of the universe requires the presence of a so called dark energy with negative pressure. Some observations indicate that the corresponding equation of state parameter is less than $-1$: $w < -1$. Such a  kind of dark energy is called ``phantom dark energy''. The evolution in the presence of such energy implies the future encounter with a cosmological singularity called ``Big Rip'' \cite{Rip,Rip1,Rip2}. 
The universe arrives in a finite interval of cosmic time to a situation wherein its scale factor and its time derivative tend to infinity. However, one can imagine a less dramatic scenario for the development of the universe, wherein the phantom or super-acceleration stage is a temporary one. In this case the universe should pass through the phantom divide line crossing which  means that the sign of the expression $w+1$ changes. Let us note that it is easy to realize  phantom dark matter by  using a ``phantom'' scalar field with 
a negative kinetic term. However, if one consider such a model, it is hardly possible to have a phantom divide line crossing. If one considers a model with two scalar fields, one normal and one phantom such a transition appears in rather  trivial way.  More interesting is the model with a unique non-minimally coupled scalar field, where one can observe the phantom divide line crossing \cite{Star-non-min0,Star-non-min}.   Another possible option is the consideration of the model, wherein there is a one scalar field with a non-analytical form of the potential \cite{ACK,CK}. On choosing properly the initial conditions in such a model one can observe the transformation of the normal scalar field into the phantom one and vice versa. Here, we wish to show that ,at least at the level  of the Friedmann model, the generalized unimodular gravity can easily describe the phantom divide line crossing.      

Indeed, let us suppose that the energy density in our universe has the form:
\begin{equation}
\varepsilon = \frac{a^2}{D+Fa^6}.
\label{cross}
\end{equation}   
In this case the pressure is 
\begin{equation}
p = -\frac53\frac{a^2}{D + Fa^6}+\frac{2Fa^8}{(D+Fa^6)^2}
\label{cross1}
\end{equation}
and 
\begin{equation}
w = \frac13-\frac{2D}{D+Fa^6}.
\label{cross2}
\end{equation}
One can then see that when 
\begin{equation}
a < \left(\frac{D}{2F}\right)^{\frac16},
\label{before}
\end{equation}
the universe is in a phantom state while for $a = \left(\frac{D}{2F}\right)^{\frac16}$ it crosses the phantom divide line and enters 
into a non-phantom phase, avoiding the future encounter with the Big Rip singularity. 

On using the formula (\ref{en-den-F}) one can find the corresponding dependence of the lapse function on the scale factor, which has a rather simple form:
\begin{equation}
N = \frac{D}{a^5} + Fa.
\label{cross3}
\end{equation} 

All said above was  concerned with the Friedmann model. It is tempting to  try to generalize these results  using the fact that in the Friedmann universe 
$\gamma = a^6$.  Then in the formulae written above it would be enough to substitute $a = \gamma^{1/6}$. However, one must  be cautious. It was shown in \cite{we} that such procedure works well for the cases when the lapse function is a power-law function of the scale factor (or of the determinant of the spatial metric), but for arbitrary functions of $a$ transformed   into  functions of $\gamma$ some difficulties can arise when we consider an arbitrary geometry of the spatial sections of our spacetime. However, it appears plausible that such a transition from the dependence on $a$ to the dependence on $\gamma$ should work for the spatially homogeneous cosmologies of the Bianchi type (see e.g. \cite{Ryan}). For this case one can represent the determinant of the spatial metric as the product of the time dependent function, obtained again as $\gamma = a^6$ and of the purely spatial part which reflects the geometry of the corresponding Bianchi manifold. (A similar trick was also 
 used in the papers \cite{Barv,Barv1}, where the auxiliary spatial metric $\sigma$ was introduced and the lapse function depended on the relation of the determinants of two metrics $\gamma/\sigma$). As far as the implementation  of  generalized unimodular gravity to  general spacetimes is concerned,
further considerations are necessary.

However, on remaining in the field of  minisuperspace models with a finite number of degrees of freedom, we can already suggest a 
further simple generalization of  unimodular gravity. Namely, the lapse function can depend not on the determinant  of the spatial metric, but on some other combination of components of the spatial part of the metric. Let us consider, for example, a hyperbolic Kantowski-Sachs universe \cite{K-S} with the metric
\begin{equation}
ds^2=N^2(t)dt^2-b^2(t)dr^2-a^2(t)(d\chi^2+\sinh^2\chi d\phi^2).
\label{K-S}
\end{equation}
The Lagrangian for the model is 
\begin{equation}
L = \frac{\dot{a}^2b}{N}+\frac{2\dot{a}\dot{b}a}{N}+Nb.
\label{K-S1}
\end{equation}
 The variation with respect to the lapse function give the constraint 
 \begin{equation}
\frac{\dot{a}^2b}{N^2}+\frac{2\dot{a}\dot{b}a}{N^2}-b = 0.
\label{K-S2}
 \end{equation} 
The variation with respect to the scale factors $a$ and $b$ gives the equations of motion:
\begin{equation}
\frac{d}{dt}\left(\frac{2\dot{a}b}{N}+\frac{2\dot{b}a}{N}\right) = \frac{2\dot{a}\dot{b}}{N},
\label{K-S3}
\end{equation}
\begin{equation}
\frac{d}{dt}\left(\frac{2\dot{a}a}{N}\right)=\frac{\dot{a}^2}{N}+N.
\label{K-S4}
\end{equation}
If we fix the time parameter by choosing the lapse function as $N=a$, we can find the metric of the Kantowski-Sachs universe in an explicit form. As one of possible solutions of Eq. (\ref{K-S4}) we have
\begin{equation}
a(t) = a_0\cosh^2\frac{t}{2}.
\label{K-S5}
\end{equation}
Then, from the constraint (\ref{K-S2}0 we find that 
\begin{equation}
b=b_0\tanh\frac{t}{2}.
\label{K-S6}
\end{equation}
It is interesting to note that there is a duality between the Kantowski-Sachs cosmological solutions and the static spherically symmetric solutions. This duality was found in paper \cite{we-dual} and further investigated in \cite{we-dual1,we-dual2}. 
If we exchange the variables $t$ and $r$ and then make the substitution $\chi \rightarrow i\theta$, we obtain the following metric \cite{we-dual}:
\begin{equation}
ds^2=b_0^2\tanh^2\frac{r}{2}dt^2-a_0^2\cosh^4\frac{r}{2}(dr^2+d\theta^2+\sin^2\theta d\phi^2).
\label{K-S7}
\end{equation}
On introducing a new variable $R \equiv a_0\cosh^2\frac{r}{2}$, we can rewrite the metric obtained  in the standard Schwarzschild form:
\begin{equation}
ds^2=b_0^2\left(1-\frac{a_0}{R}\right)dt^2 - \frac{dR^2}{1-\frac{a_0}{R}}-R^2(d\theta^2+\sin^2\theta d\phi^2).
\label{K-S8} 
\end{equation}
Let us now suppose that we at the beginning have fixed $N=a$. We note that in this case the lapse function is not a function of the determinant of the spatial metric, thus we consider a further generalization of  unimodular gravity. The constraint \eqref{K-S2} then disappears and the equations of motion \eqref{K-S3}-\eqref{K-S4} are modified, because their right-hand sides should include the terms proportional to the partial derivatives of $N$ with respect to $a$ and $b$.  On multiplying these modified equations of motion by $\dot{a}$ and $\dot{b}$ respectively, one can see that their sum gives the total derivative of the new first integral which is
\begin{equation}
 \frac{\dot{a}^2b}{N}+\frac{2\dot{a}\dot{b}a}{N}-Nb =A,
 \label{K-S9}
 \end{equation}
 where $A$ is an arbitrary constant. We thus obtain our  constraint, where  an effective matter arises  with the energy density
 equal to $\varepsilon = \frac{A}{a^3b}$. Let us find the solutions to the field equations of this new theory.  Firstly, we note that Eq. \eqref{K-S4} has not changed because $N$ does not depend on $b$. Thus the expression for $a$ is the same as before. Now, on using Eq. \eqref{K-S9}, we find that the scale factor $b$ is 
 \begin{equation}
 b = b_0 \tanh\frac{t}{2}-\frac{A}{a_0}.
 \label{K-S10}
 \end{equation}
On using the same duality relations and changes of variable, we obtain the following Schwarzschild-type metric:
\begin{eqnarray}
&&ds^2=\left[b_0^2\left(1-\frac{a_0}{R}\right)-2\frac{Ab_0}{a_0}\sqrt{1-\frac{a_0}{R}}+\frac{A^2}{a_0^2}\right]dt^2 \nonumber \\
&&- \frac{dR^2}{1-\frac{a_0}{R}}-R^2(d\theta^2+\sin^2\theta d\phi^2).
\label{K-S11}
\end{eqnarray}
We see that while the spatial part of the metric has not changed, the coefficient $g_{00}$ at $dt^2$ has changed essentially. If the constant $A$ 
is positive then  the metric coefficient vanishes at 
$$
R_0 = \frac{a_0}{1-\frac{A^2}{a_0^2b_0^2}} > a_0,
$$ 
provided $ A^2 < a_0^2b_0^2$.  
We should then think how to describe the continuation  of the metric in the region where $R < R_0$ and then to $R< a_0$.
If $A$ is negative (the energy density of the effective matter is negative) the expression for $b$ cannot become equal to zero, but we still stumble upon the problem of its behaviour fo $R < a_0$. 

On summing up, we wish to say that  unimodular gravity theory can be treated as a theory wherein part of its symmetries  are frozen (the action remains invariant with respect to the subgroup volume-preserving diffeomorphisms instead of the whole group 
of diffeomorphisms). This leads one to consider its generalization by treating the lapse function as function of the determinant of the spatial part of the metric \cite{we}. Thus, the number of constraints decreases and  some effective matter emerges. We have illustrated some features of this approach on the simplest example of a Friedmann universe. A more complicated example of a Kantowski-Sachs universe suggests a further generalization of this scheme, wherein the lapse function depends on the spatial metric in a more complicated way. One then observes a general phenomenon: on decreasing the symmetry of the system a more complicated theory with some additional degrees of freedom and some new effects is obtained.  
We feel that the mathematical structure and the possible physical effects arising in the context  of  generalized unimodular gravity 
and similar models are worthy of further investigation. 

We are grateful to A. O. Barvinsky and T. Vardanyan for fruitful discussions. The work of Alexander Kamenshchik was partially supported by the Russian Foundation for Basic Research   grant  No. 18-52-45016.

\end{document}